\documentclass[a4paper,10pt,{twoside}]{cpc-hepnp}

\usepackage{multicol}
\usepackage{graphicx}
\usepackage{epstopdf}
\usepackage{booktabs}
\usepackage{amssymb,bm,mathrsfs,bbm,amscd}
\usepackage[tbtags]{amsmath}
\usepackage{lastpage}

\def\p{\partial}

\def\g{\gamma}

\def\de{\delta}
\def\D{\Delta}
\def\De{\Delta}
\def\ov{\overline}
\def\ld{\lambda}
\def\Ld{\Lambda}

\def\th{\theta}

\def\om{\omega}

\def\rh{\rho}

\def\pdellx'{\frac{\partial}{\partial x'}}
\def\pdellw'{\frac{\partial}{\partial w'}}

\newcommand{\be}{\begin{equation}}
\newcommand{\ee}{\end{equation}}
\def\bed{\begin{displaymath}}
\def\eed{\end{displaymath}}
\def\bea{\begin{eqnarray}}
\def\eea{\end{eqncrray}}

\begin{document}

\fancyhead[c]{\small Chinese Physics C~~~Vol. 41, No. 1 (2017) 015101}
\fancyfoot[C]{\small 010201-\thepage}



\title{Quark Confinement, New Cosmic Expansion and \\ General Yang-Mills Symmetry}

\author{%
Jong-Ping Hsu$^{1;1)}$\email{jhsu@umassd.edu}%
}
\maketitle

\address{%
$^1$ Department of Physics, University of Massachusetts Dartmouth, North Dartmouth, MA 02747-2300, USA\\
}

\begin{abstract}
We discuss a unified model of quark confinement and new cosmic expansion with linear potentials based on a general  $(SU_3)_{color} \times (U_1)_{baryon}$ symmetry.  The phase functions in the usual gauge transformations are generalized to new `action integrals'.  The general Yang-Mills transformations have group properties and reduce to usual gauge transformations in special cases.  Both quarks and `gauge bosons' are permanently confined by linear potentials.  In this unified model of particle-cosmology, physics in the largest cosmos  
 and that in the smallest quark system appear to  both be dictated by the general Yang-Mills symmetry and characterized by a universal length.  The basic force between two baryons is independent of distance. However, the cosmic repulsive force exerted on a baryonic supernova by a uniform sphere of galaxies is proportional to the distance from the center of the sphere.  The new general Yang-Mills field  may give a field-theoretic explanation of the accelerated cosmic expansion.  The prediction could be tested experimentally by measuring the frequency shifts of supernovae at different distances.
 \end{abstract}



\footnotetext[0]{\hspace*{-3mm}\raisebox{0.3ex}{$\scriptstyle\copyright$}2017
Chinese Physical Society and the Institute of High Energy Physics
of the Chinese Academy of Sciences and the Institute
of Modern Physics of the Chinese Academy of Sciences and IOP Publishing Ltd}%

\begin{multicols}{2}

\section{Introduction}

We discuss a unified model with a general Yang-Mills symmetry with the groups,  $(SU_3)_{color} \times (U_{1})_{baryon} \equiv SU_{3c} \times U_{1b} $.   The model is consistent with all established conservation laws, including the conservations of color charge and baryon charge (or number)~\cite{1,2}.\footnote{The leptonic $U_1$ group is not considered here for simplicity.  It can be included in the model without difficulty.  See Ref. [4].}  The general Yang-Mills symmetries for particle-cosmology~\cite{3} are based on generalized gauge transformations, in which the usual scalar gauge functions $\om^a$ and phase factors are replaced by vector gauge functions $\om^a_\mu$ and non-integrable phase factors.  The non-integrable phase is an action integral, which involves a fixed initial point and a variable end point.  In previous works~\cite{4,5}, the generalized gauge transformations with the new phase functional was not unambiguously specified.  As we shall discuss below, for this new action integral to have an unambiguous partial derivative, one has to impose a Lagrange equation to specify the path, similar to Hamilton's characteristic function\cite{6}, which is a local function.  Therefore, the results of the general Yang-Mills transformations do not lead to non-local gauge fields.   

Moreover, the general Yang-Mills symmetry leads to a fourth-order equation for massless gauge fields (or `confions'), which could provide an explicit mechanism to confine quarks and massless confions, and to understand quark energy spectra and others.  The reason is as follows:  The new fourth-order  gauge field equations have unusual source terms, which suggest two possible types of sources in the static limit. One is the usual delta function source $\de^3({\bf r})$, and the other involves the derivative of the delta function $ \D \de^3({\bf r})$. These sources lead to a dual potential (i.e., a linear and a Coulomb-type potentials) together with a universal scale length $L_s$ associated with the gauge invariant Lagrangian.  Thus, the new gauge symmetry provides a  mechanism with dual potentials to confine quarks.
The fourth-order field equation is usually considered unphysical because the dynamical system involves non-definite energy~\cite{7}.  In other words, the gauge bosons associated with the fourth-order field equation have negative energies.  However, in the present model, the gauge bosons are permanently confined in the quark system and, hence, do not have observable negative energies. 
Nevertheless, other new ideas are needed for a satisfactory S matrix with the general Yang-Mills symmetry.   For example, the model also suggests that these new gauge bosons should be off-mass-shell particles.  These properties, including quantization of general Yang-Mills fields and, furthermore, physical implications of the unified model, will be discussed in a separate paper.

For physics at the cosmic scale, a basic question is whether we can have a field-theoretic understanding of the late-time accelerated cosmic expansion.  The answer turns out to be affirmative.  The unified $SU_{3c} \times U_{1b} $ model for particle cosmology is consistent with a Big Jets model of the universe~\cite{4}, which is consistent with all established conservation laws, including an  equal amount of matter and anti-matter in the physical universe, in contrast to the usual Big Bang model.  The two big jets may be pictured as two gigantic fireballs moving away from each other.  The evolution of each fireball could be similar to a big bang.  Presumably, beyond what we can detect now, there may be a gigantic cluster of anti-galaxies dominated by anti-baryons (and anti-leptons).   In the unified  model, the baryonic gauge fields predict that there is an extremely weak repulsive force independent of distance between two baryons.  However, the more realistic cosmic force between a big uniform sphere of galaxies and a supernova is shown to be a linear force, called Okubo force.  This differs from that of the gravitational inverse-square force.  Such a cosmic baryon force will be stronger than the gravitational force at sufficiently large distances.  Therefore, the dominant repulsive force between two baryon galaxies at extremely large distances in the unified model can play the role of `dark energy' for late-time accelerated cosmic expansion.  These properties and their  possible experimental test are discussed below.

\section{General Yang-Mills symmetry}

(A) General $U_1$ symmetry

To construct a Lagrangian with the general Yang-Mills (gYM) symmetry, we consider the new gauge transformations.  The new $U_{1b}$ gauge transformations for quarks $q(x)$ are
\be
\psi'(x) = e^{-iP_1}\psi(x), \ \ \ \   \ov{\psi}'(x) = \ov{\psi}(x) e^{+iP_1},  \ \ \ \   \psi(x)=q(x),
\ee
where $P_1$ is an `action integral' with a variable end point $x'_e =x$ and a fixed initial point $x'_o$
\be
  P_1=\left( g_b \int_{x'_o}^{x'_e =x} dx'^\mu \Ld_\mu(x')\right)_{Le1}, \ \ \ \  c=\hbar=1,
\ee
which resembles Hamilton's characteristic function\cite{6}.  New gauge transformations for the $U_{1b}$ gauge fields, $B_\mu(x)$, are given by
 \be
 B'_\mu(x) =B_\mu(x) + \Ld_\mu(x),
 \ee
which involves a vector gauge function $\Ld_\mu(x)$ rather than the usual scalar function.  To see the meaning of the subscript  $Le1$ in (2), let us consider the variation of $P_1$.  We have
$$
 \de P_1 =g_b  \Ld_\mu(x')\de x'^{\mu} |_{{x'_o}}^{x'_{e}=x} 
 $$
 \be
 +g_b \int_{x'_o}^{x}\left(\frac{\p\Ld_\mu(x')}{\p x'^\ld} dx'^\mu - d\Ld_{\ld}(x')\right)\de x'^\ld.
 \ee

We can determine the variation of the action functional $P_1$ as a function of the coordinates because the initial point $x_o$ is fixed, so that $\de x'(x'_o)=0$ and the end point is variable, $x'_e=x$.  Furthermore, the path in (4) is required to satisfy~\cite{8} the Lagrange equation $Le1$, i.e., 
\be
 d\Ld_{\ld}(x')-\frac{\p\Ld_\mu(x')}{\p x'^\ld} dx'^\mu=0, \ \ \  
 \ee
 $$
 or  \ \ \  (\p_\mu \Ld_\ld(x)-\p_\ld \Ld_\mu(x))dx^\mu=0,
 $$
where $x'=x$.  
This requirement is necessary for the integral in (4) to vanish.  As a result, we have
\be
\de P_1 =g_b \Ld_\mu(x)\de x^\mu, \ \ \ \ \  or  \ \ \ \  \p_{\mu} P_1=g_b \Ld_\mu(x).
\ee
 This equation for the $U_{1b}$ phase `characteristic function' $P_1$ is crucial for a Lagrangian to have the general Yang-Mills symmetry. 
  
 One can verify that $\ov{\psi}(\p_\mu + ig_b)\psi$ is invariant under the general $U_{1b}$ transformations (1) and (3),
 \be
\ov{\psi}'(\p_\mu + ig_b B'_\mu)\psi' = \ov{\psi}(\p_\mu + ig_b B_\mu)\psi, 
\ee
where we have used equations (1), (2), (3) and (6).  

The general  $U_{1b}$ gauge covariant derivative, $\p_\mu + ig_b B_\mu$, in (7) is the same as that in the usual $U_1$ gauge theory QED.
The $U_{1b}$ gauge curvature is, as usual, determined by the commutator of the covariant derivative,
\be
[\p_\mu + ig_b B_\mu, \p_\nu + ig_b B_\nu]= ig_b B_{\mu\nu},
\ee
 $$
 B_{\mu\nu}=\p_\mu B_\nu - \p_\nu B_\mu.
$$
Under the general $U_{1b}$ transformation (3), the $U_{1b}$ gauge curvature $B_{\mu\nu}$ turns out not to be invariant,
$
 B'_{\mu\nu}=B_{\mu\nu}+\p_\mu \Ld_\nu - \p_\nu \Ld_\mu \ne B_{\mu\nu} .
$
However, the divergence of the gauge curvature, $ \p^\mu B_{\mu\nu}$,  can be invariant,
\be
\p^\mu B'_{\mu\nu}= \p^\mu B_{\mu\nu},
\ee
provided the vector functions $\Ld_\mu(x)$ satisfy the constraint equations,
\be
\p^\mu[\p_\mu \Ld_\nu(x) -\p_\nu \Ld_\mu(x)]=0.
\ee

In special cases, if the vector function $\Ld_\mu(x)$ can be expressed as the space-time derivative of an arbitrary scalar function $\Ld(x)$, i.e., $\Ld_{\mu}=\p_{\mu} \Ld(x)$, the Lagrange equation (5) and the constraint (10) become identities for arbitrary function $\Ld(x)$ and the transformations (1), (2) and (3) become the same as the usual $U_1$ gauge transformations.   However, in the general $U_1$ transformations (1) and (3), the vector function $\Ld_\mu(x)$ does not take the special form $\p_{\mu} \Ld(x)$.  
 \bigskip

\noindent 
(B) General $U_1$ group properties

 For the general $U_1$ transformations (1) and (3), the group operations have inverse, identity (i.e.,  $\Ld_\mu=0$)  and associativity.  To see closure, let us consider two consecutive transformations $exp(-iP_a)$ 
 and $exp(-iP_b)$, we have  
 \be
exp(-i[P_a+P_b])= exp(-iP_c), \ \ \ \ \  
\ee
 $$
 P_c=\left( g_b \int_{x'_o}^{x'_e =x} dx'^\mu \left[\frac{}{}\Ld_{c\mu} (x') \right]\right)_{Le{c}}
 $$
where  $\Ld_{c\mu}=\Ld_{a\mu} +\Ld_{b\mu} $ and  the Lagrange equation $L_{ec}$ is given by $[\p_\mu \Ld_{c\ld}(x)-\p_\ld \Ld_{c\mu}(x)]dx^\mu=0 $.
One can verify that the results in (4)-(10) are still true for the consecutive transformations, where $P_1$ and $\Ld_\mu$ are replaced by  $P_c$ and $\Ld_{c\mu}$.  Thus, the general transformations (1), (2) and (3) together with the constraint (10) can be considered as a generalization of $U_1$ gauge transformations.
\bigskip

\noindent
(C) General $SU_N$ symmetry

The discussions of the general $U_{1b}$ group can be applied to the non-Abelian group $SU_N$, which includes $SU_{3c} $ as a special case. The general $SU_N$ transformations for quarks $q(x)$ are given by
\be
q'(x) = (1- iP)q(x), \ \ \ \   \ov{q}'(x) = \ov{q}(x) (1+iP),  
\ee
where $P$ is an infinitesimal phase `characteristic function' (or `action integral'), 
\be
  P=L^a\left( g_s \int_{x'_o}^x dx'^\mu \om_\mu^a (x') \right)_{Le} =L^a P^a(\om,x),
\ee
\be
[L^a, L^b]= if^{abc} L^c,
\ee
where $\om=\om^a_\mu$ are arbitrary vector functions and $L^a$ are $SU_N$ generators\cite{9}.
 The general Yang-Mills (gYM) transformations for the $SU_{N}$ gauge fields, $H^a_\mu(x)$ are given by
 \be
 H'_\mu(x) =H_\mu(x) + \om_\mu(x)  -i [P(x), H_{\mu}(x)], \ \ \ \ \    H_{\mu}=H^a_{\mu}L^a.
  \ee
  or
  \be
   H'^a_\mu(x) =H^a_\mu(x) + \om^a_\mu(x) + g_s f^{abc} P^b(\om,x) H^c_{\mu}(x),
   \ee
 where $P^b(\om,x)$ is defined in (13).  
To see the equation  $Le$ in (13), let us consider the variation of $P$ with a variable end point $x$ and a fixed initial point $x'_o$.  We have
\be
 \de P =g_s  \frac{\p L}{\p\dot{x}^\ld}\de x^{\mu}  + g_s \left( \int_{\tau_o}^{\tau}\left(-\frac{d}{d\tau}\frac{\p L}{\p \dot{x}^\ld}  
 + \frac{\p L}{\p x^\ld}\right)\de x^\ld d\tau\right)_{Le}
\ee
where we write (13) in the usual form of a Lagrangian with the help of a parameter $\tau$,
\be
P=\left( g_s \int_{\tau_o}^\tau L  \ d\tau\right)_{Le}, \ \ \    L=\dot{x}^\mu \om_\mu^a (x) L^a, \ \ \  \dot{x}^\mu=\frac{dx^\mu}{d\tau}.
\ee

We require that the paths in (17) are those that satisfy the Lagrange equation $Le$, i.e., 
\be
-\frac{d}{d\tau}\frac{\p L}{\p \dot{x}^\ld} + \frac{\p L}{\p x^\ld}=0,    \ \ \  L=\dot{x}^\mu \om_\mu^a (x) L^a.
\ee
This Lagrange equation can also be written in the form 
\be
\frac{d \om^a_{\ld}}{d \tau} - \frac{\p \om^a_{\mu}}{\p x^\ld} \dot{x}^{\mu} = 0,
\ee
similar to equation (5) for the general $U_{1b}$ symmetry.
Thus, the integral in (17) vanishes and we have the relation\cite{8}
\be
 \p_{\mu} P=g_s\om^a_\mu(x) L^a = g_s\om_\mu.
\ee
 
In the following discussion, we shall concentrate on the specific case N=3.  As usual, the $SU_{3c} $ gauge covariant derivatives are  defined as
\be
 \De_{\mu} = \p_{\mu} + ig_{s}{H_{\mu}^{ a}}\frac{{ \ld^a}}{2},   \ \ \ \  L^a= \frac{{ \ld^a}}{2},
\ee
where $\ld^a$ are the $3 \times 3$ Gell-Mann matrices.\cite{9}
The $SU_{3c}$ gauge curvatures $H^a_{\mu\nu}$ are given by
\be
[\De_\mu, \De_\nu]= ig_s H_{\mu\nu}, 
\ee
\be
H_{\mu\nu}=\p_\mu H_\nu - \p_\nu H_\mu + ig_s[H_\mu, H_\nu],
\ee
or
$$
H^a_{\mu\nu}=\p_\mu H^a_\nu - \p_\nu H^a_\mu - g_s f^{abc}H^b_\mu H^c_\nu.
$$

It follows from equations (15)-(24) that we have the following gYM transformations for $\p^\mu H_{\mu\nu}(x)$, and $\ov{q}\De_{b\mu} q$:
\be
\p^\mu H'_{\mu\nu}(x)= \p^\mu H_{\mu\nu}(x)- [P(x), \p^\mu H_{\mu\nu}(x)]
\ee
\be
 \ov{q}' \g^\mu \De'_{\mu} q'  =  \ov{q}\g^\mu  \De_{\mu} q  , 
 \ee
provided the restrictions  
\be
\p^\mu \{\p_\mu \om_\nu (x) - \p_\nu \om_\mu (x)\} + ig_s [\om^\mu (x), H_{\mu\nu}(x)] = 0
\ee
are imposed for (25) to hold.  This constraint is similar to that for gauge functions of Lie groups in the usual non-Abelian gauge theories\cite{10}.

\bigskip

\noindent
(D) General $SU_N$ group properties

 For group properties of the general $SU_N$ transformations (12) and (15), it is convenient to consider infinitesimal transformation, $e^{-iP}\approx (1-iP)$.  Clearly, we have inverse, identity and associativity for group operations.  To see the closure property,  we consider two consecutive transformations involving  $ \om^a_{r\mu}$ and $\om^a_{s\mu}$.  We have 
$$
(1-iP_r)(1-iP_s)= (1-iP_t),   \ \ \ \ \   P_t=P_r + P_s
$$
\be
 P_t=\left( g_s \int_{x'_o}^{x'_e =x} dx'^\mu \left[\om^a_{t\mu} (x') L^a \right]\right)_{Le(t)}.
\ee
$$
\om^a_{t\mu} (x')L^a =\om^a_{r\mu} (x')L^a + \om^a_{s\mu} (x')L^a. 
$$
The Lagrange equation $Le(t)$ in (28) is given by (19) with $L= \dot{x}^\mu \om^a_{t\mu}(x) L^a$.

After some tedious but straightforward calculations, the results in (17)-(21) are still true for two consecutive transformations, where $P$ and $\om^a_\mu L^a$ are respectively  replaced by  $P_t$ and $\om^a_{r\mu}L^a+\om^b_{s\mu}L^b$.   Furthermore, the result (25) with $P(x)$ replaced by $P_t$ holds, provided we impose the constraint (27) with $\om_\mu(x)$ replaced by $\om_{r\mu}(x)+\om_{s\mu}(x)$.
Thus, the general transformations (12)-(15) together with the constraint (27) could be considered as  general $SU_N$  transformations with the required group properties.   (See Appendix).

\section{A unified model for quark confining potential and cosmic repulsive force}

Let us construct a unified model with the invariant Lagrangian based on the general Yang-Mills symmetry involving the gauge groups  $[SU_{3c}\times U_{1b}]$.  The general Yang-Mills (gYM) invariant Lagrangian $L_{gYM}$  is
\be
L_{gYM} = \frac{L_s^2}{2}[ \p^\mu H^a_{\mu\ld} \p_\nu H^{a\nu\ld} +  \p^\mu B_{\mu\ld} \p_\nu B^{\nu\ld}] 
\ee
$$
+i \ov{q}[\g^\mu(\p_{\mu} - ig_{s}{H_{\mu}^{ a}}\frac{{ \ld^a}}{2} - ig_{b}{B_{\mu}})  - M]q.
$$
  The matter fields consist of spinor quark fields q(x) with components $q^{fi}$, where $i=1,2,3$ for color indices and $f=1,2,....6$ for flavor indices (u,d,c,s,t,b)  and $M$ is a color-independent mass matrix in the flavor indices\cite{9}. 
  
  The length $L_s$ denotes a universal scale with the dimension of length, which characterizes the dynamics of the cosmos and quark systems described by the gYM Lagrangian (29).  It plays a dual role in particle cosmology as follows: 
  
(i) $L_s$ is the universal and fundamental length for all gYM fields in (29).   Its value can be determined from the data of quark spectra.

(ii) The fundamental length $L_s$, together with the baryon coupling constant $g_b$, determines the strength of the cosmic repulsive force between, say, a supernovae and a uniform sphere of baryon galaxies.  

From the gYM Lagrangian (29), one can derive field equations for $H^a_\mu$ and quarks,
\be
\p^2 \p^\mu H^a_{\mu\nu} - (g_s/L^2_s)\ov{q} \g_\nu (\ld^a/2) q = 0,
\ee
and 
\be
(i\g^\mu [\p_{\mu} - ig_{s}{H_{\mu}^{ a}}\frac{{ \ld^a}}{2} - ig_{b}{B_{\mu}}] - M) q = 0. 
\ee

The sources of gYM fields in (30) include the usual quark source and the self-coupling of `confions,' i.e., $ g_s f^{abc}(\p_\ld \p^{\mu} H^b_{\mu\nu} -\p_{\nu}\p^{\mu} H^b_{\mu\ld}) H^{\ld c}$.  In the static limit, these sources for the time-component $H_0$ of the gYM field (with a given index $a$) are interpreted to be
$j_0 = j_{0\psi} + j_{0 H} \approx -g_s \de^3({\bf r}) + g_s L_s^2\nabla^2 \de^3({\bf r})$.  This result appears to be consistent with the dimensional analysis of source terms in (30).  Thus, the static solution of $L_s^2 \nabla^2 \nabla^2 H_0 = j_0$ is\cite{5}
\be
H_0 =  g_s [ r/(8\pi L_s^2) - 1/(4\pi r)].
\ee
$$
\frac{g^2_s}{4\pi} \approx 0.04, \ \ \ \ \ \ \  L_s \approx 0.28 fm,
$$
where the potential $g_s/(4\pi r)$ is produced by the source $j_{0H}$.  The quanta associated with the new gYM fields $H_\mu$ may be called `confions.'  The dual quark potentials in (32) are consistent with and supported by the empirical charmonium potential, which was obtained by the Cornell group by fitting the charmonium spectrum with the charmed quark mass $m_c \approx 1.6 GeV$\cite{11}.

The equations of the gYM field $B_\mu$ are 
\be
L_{s}^2 \p^2 \p^\ld B_{\ld\mu} -  g_b\ov{q} \g_\mu q = 0,
\ee
where the $B_\mu$ fields do not have self-coupling and are generated by quarks, which carry baryon charges $g_b$ in addition to color charge $g_s$. If we choose a gauge condition $\p^\ld B_\ld =0$, (33) leads to the following field equations
\be
\p^2 \p^2 B_\mu   =  \frac{g_b}{L^2_s} \ov{q}\g_\mu q.
\ee
Suppose one puts a point-like baryon charge at the  origin.   The zeroth component static gauge potential $B_0({\bf r})$ satisfies the fourth-order equation,
\be
\nabla^2 \nabla^2 B_0 ({\bf r})=  \frac{g_b}{L^2_s} \de^3({\bf r}).
\ee
In this case, we have only the usual point charge in (35) due to the quark source in (34).
It leads to a linear gauge potential $B_0({\bf r})$\cite{2,5} and the repulsive  cosmic force $F_{bb}$ between two baryon charges $g_b$
\be
B_0({\bf r}) =- \frac{g_b}{L^2_s}\frac{ r}{8\pi}, \ \ \ \ \ \ \ \ \  {\bf F}_{bb}=- g_b\boldsymbol{\nabla} B_0=\frac{1}{8\pi}\frac{g^2_b}{L^2_s}\frac{{\bf r}}{r},
\ee
where we have used the relation for generalized functions,\cite{12}   
\be
\int^{\infty}_{\infty}\frac{1}{({\bf k}^2)^2} exp(i{\bf k .  r}) d^3 k = -\pi^{2} r.
\ee

\section{Cosmic Okubo force for a uniform \\ sphere of galaxies and a supernovae}

The cosmic force between two baryons is a constant (or distance-independent) force rather than an inverse-square force.  Thus, one cannot simply assume that the entire baryon charge in a uniform sphere acts as if concentrated at its center and produces a distance-independent force\cite{13}.

\begin{center} 
\hspace*{-0.9cm}
\includegraphics[width=12cm]{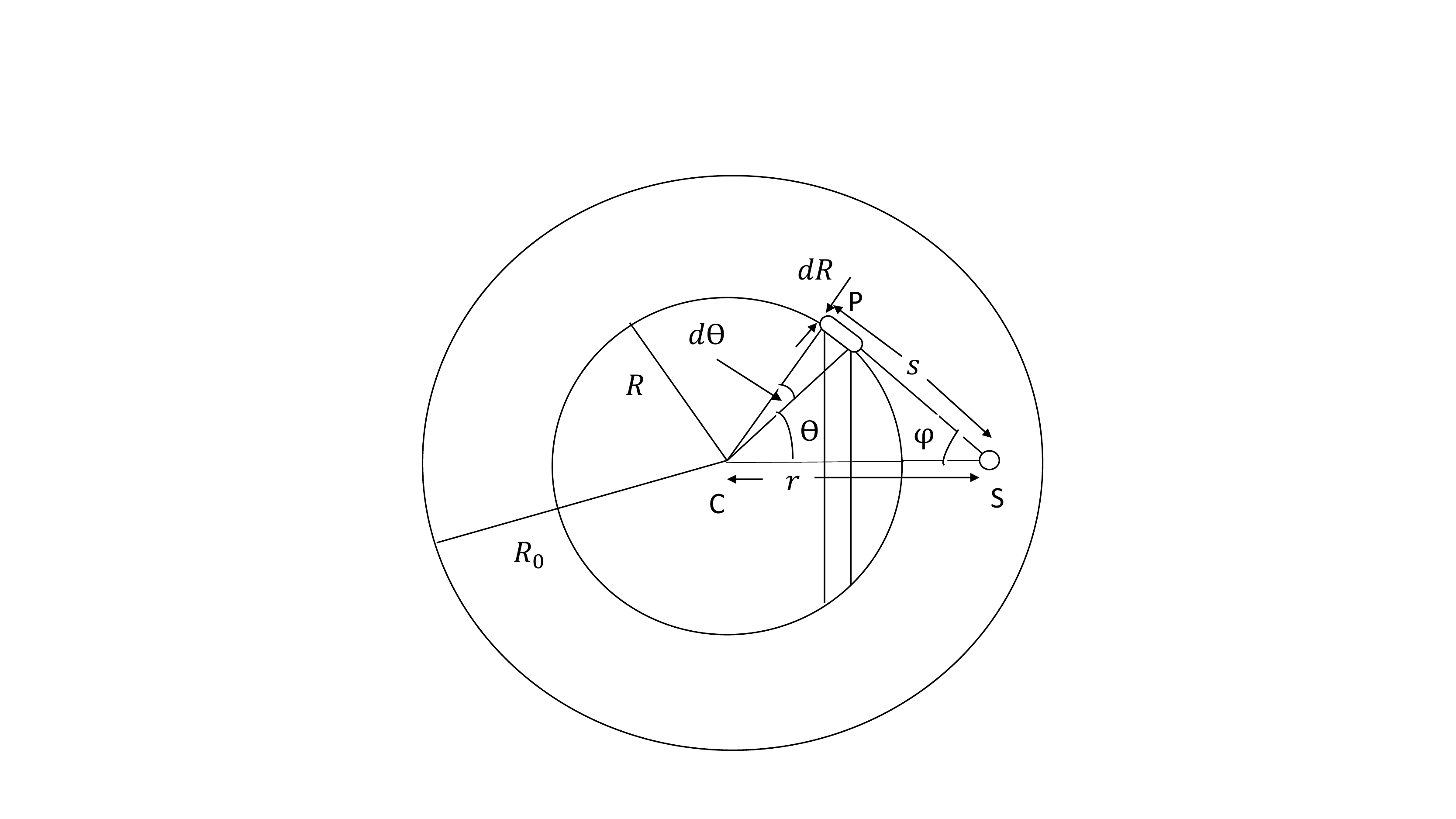}
\figcaption{     A schematic diagram for calculations of the cosmic baryon (Okubo) force $F_{CS1}$ between a gigantic uniform sphere of baryon galaxies and a point-like baryon supernova.}

\end{center}


Let us calculate the cosmic force between a gigantic uniform sphere (with a radius $R_o$) of baryon galaxies and a baryon supernova, which is idealized as a point with mass m or with baryon charge $(3g_b)m/m_p$, where $m_p$ is the proton mass and each proton (or neutron) carries three baryon charges, $3g_b$.  We consider a thin uniform shell of radius $R<R_o$\cite{13}. The density of baryon mass is $\rh$ on the shell with a thickness $dR$.  Suppose the distance of the center $C$ of the sphere to the supernova at $S$ is denoted by $r$,  $r>R$.  Suppose a point $P$ is on a ring, which is perpendicular to $CS$, and the angle $SCP$ is denoted by $\th$.  The mass $dM_r$ of the ring is $dM_r = \rh 2\pi R^2 dR \ sin\th d\th,$ where $2\pi R sin\th$ is the circumference of the ring.  The repulsive baryon force $dF_P$ exerted on the supernova at $S$ by a small sub-element $T$ in the direction PS.  Suppose the angle CSP is denoted by $\phi$. The force $dF_P$ can be decomposed into a component $dF_P cos\phi$ along $CS$ and many perpendicular components $dF_P sin\phi$. These perpendicular components cancel due to symmetry and do not contribute from the entire ring\cite{13}.

Let us calculate the force $dF_{CS}$ exerted on the supernova $S$ by the entire ring in the direction $CS$.  Using (36), we have
\be
dF_{CS}=\int\frac{(3g_b)^2 m (dM_r)}{8\pi L^2_s m_p^2} cos\phi ,   \ \ \      dM_r=(\rh 2\pi R^2 sin\th d\th)  dR,
\ee
where $dM_r/m_p$ is the number of baryons (with mass $m_p$) in the ring and each baryon carries three baryon charge, $3g_b$. 
We have the relations $s^2=R^2+r^2 -2Rr \ cos\th$ and $R^2=s^2 +r^2 -2rs \ cos\phi$, where $s$ is the distance $PS$.  The integration of $d\th$ from $0$ to $\pi$ leads to an effective repulsive force on $S$ in one direction.  Thus,  the resultant force is given by
\be
dF_{CS} =A\int _{0}^{\pi} \left[dR \ R^2 cos\phi\right] sin\th \ d\th = A \int_{R-r}^{R+r} [dF ]\frac{s }{Rr}ds,
\ee
$$
 A=\frac{(3g_b)^2 m \rh}{4 L^2_s m_p^2},  \ \ \  dF \equiv dR \ R^2\left[ \frac{s^2+r^2-R^2}{2sr}\right].
$$
The magnitude of the total repulsive force  can be obtained~\cite{13} from the integration of $ds$ from $r-R$ to $r+R$ with $r>R$ and that from  $R'-r$ to $R'+r$ with $R_o > R' >r$, 
$$
F_{CS1} =A\int _{0}^{r} \frac{R dR}{2r^2}\int_{r-R}^{r+R} ds (s^2+r^2 -R^2) 
$$
$$
F_{CS2} =A\int _{r}^{R_o} \frac{R' dR'}{2r^2}\int_{R'-r}^{R'+r} ds (s^2+r^2 -R'^2) 
$$
In the calculation of $F_{CS2}$, we modify Fig.~1 such that the distance CS is smaller than $R'(=R)$.  The magnitude of the cosmic baryon  force $F_{cbf}=F_{CS1} + F_{CS2} $ of the gigantic sphere with baryon galaxies exerted on a point-like supernova S is 
\be
F_{cbf} = \left(\frac{(3g_b)^2 m M}{8\pi L^2_s m^2_p}\right)\left[\frac{r}{R_o}-\frac{r^3}{5 R_o^3}\right],  \ \ \ \ M=\frac{4\pi R_o^3\rh}{3},
\ee
which  turns out to be dominated by a linear force for $r << R_o$.   We shall call this new cosmic  baryon force based on gYM symmetry `Okubo force'  in memory of his endeavor.
In this calculation, we have approximated the observable portion of the universe by a gigantic uniform sphere with a finite radius $R_o$.\footnote{A simple model corresponds to a finite $R_o$.  If one wishes, one may take the limit $R_o \to \infty$, provided that the density $\rh $ also approaches zero such that $R^2_o\rh$ remains finite.}

\section{Discussion}  

 We stress that the group properties in (11) for $U_{1b}$ and  those in (28) for $SU_{3c}$ are useful for the renormalization of the unified model based on the Lagrangian $L_{gYM}$ in (29) for particle cosmology.  The renormalization of quark interactions is particularly important because the usual results in quantum chromodynamics will not be upset by the higher order corrections based on the Lagrangian (29) with general Yang-Mills symmetry~\cite{14,15}.

Let us consider the cosmological implications of the $U_{1b}$ linear Okubo force (40), produced by the baryonic gYM field at a cosmic scale.  The distance between baryon galaxies was much smaller in an earlier era and the gravitational attractive force dominated the motion of matter.  In this era, we expect to have a decelerated cosmic expansion.  As the universe expanded, the inverse-square gravitational force decreases.  There is a critical distance $R_c$ where the repulsive Okubo force (40) cancels the gravitational attractive force between the uniform sphere and a supernova.  In this case, we could relate the length scales $R_c$, $R_o$ and $L_s$ by the relation,
\be
R_c^3 \approx \left(\frac{G m_p^2 8 \pi L_s^2 R_o}{9 g_b^2} \right). 
\ee
Although there is no data for $R_c$ and $R_o$, they are presumably the same order of magnitude as the size of the observable universe.  To get a feeling for the order of magnitude of the baryon charge, let us use a `reasonably' large value  $R_c\approx  10^{22} m\approx R_o$ and use the basic scale length $L_s$ in (32) to roughly estimate the coupling constant $g_b$.  We obtain
\be 
 g_b \approx 10^{-50}, 
 \ee
 For a very wide range of cosmic lengths $R_c$ and $R_o$, the dimensionless baryon charge $g_b$ and the Okubo force are extremely small.  However, it can have observable effects in the cosmic arena involving extremely large numbers of baryons, such as those involved in the accelerated cosmic expansion.  Such an extremely small Okubo force in (40), the baryonic charge $g_b$ in (42) and the associated massless boson (or gYM fields $B_\mu$) simply cannot be detected in the high energy laboratory or in our solar system, or in the Milky Way galaxy.  
 
The nature of color SU(3) in particle physics does not allow regular color charge (or the strong force) to play a role in the accelerated cosmic expansion.  The reasons are as follows. The SU(3) Yang-Mills fields and quarks with color charges are confined.  This implies color states do not exist in the physical sector of Hilbert space.  In other words, the basis states of the S matrix exclude a single quark or `confion' state in this model.  Physical hadrons in stars and galaxies are color singlets.   In the present model, only the baryonic (and the leptonic) U(1) sectors play a role in the accelerated comic expansion.

The U(1) sector  in the unified model helps indirectly to suggest a new view of the baryon asymmetry in the observable Universe.   All gauge invariant interactions in the unified model appear to dictate particle-antiparticle symmetry, so that particles and antiparticles must have been created  in equal numbers in the Universe.  This symmetry property should also hold  for  annihilation and decay processes in general.  To be consistent with all these properties, we suggest a new model that the universe began with two Big Jets\cite{4,16,17} rather than one Big Bang, similar to the type of phenomena one might encounter in particle collisions in high-energy laboratories.  According to this model, the universe originated with the formation of Big Jets, i.e., two diametrically opposed jets, composed of baryons and anti-baryons, etc. in each jet.   We may picture these two jets as two big fireballs moving away from each other.  The processes of their annihilations eventually lead to a baryon dominated fireball and an anti-baryon dominated fireball.  Then, from the vantage point of an observer in either fireball, the evolution of that `observer's universe' would be similar to the general features of a Big Bang.  A detailed description of specific phenomenon implied by a Big Jets model awaits future investigation. 

 There are many discussions for the physical origin of the accelerated cosmic expansion.  Apart from the cosmological constant in Einstein's equation, there are discussions of a dark energy model based on a non-perturbative Yang-Mills condensate~\cite{18}.  In this approach, one uses an effect action with certain desirable properties to drive the universe toward a cosmological dark energy phase.  So far, it has not led to definite results that can be tested experimentally.
 
Let us consider the dark matter in a baryon galaxy, which moves under the repulsive Okubo force.  Although we do not know what dark matter particles are, we may consider a very simple model for discussion and assume that they are stable and have only weak and gravitational interactions, without contradicting experiments and observations~\cite{19}.  Since the weak forces produced by the exchanges of $W^{\pm}$ and $Z^0$ bosons are very short range forces, they do not affect the macroscopic motion of dark matter particles with low density.  Let us consider two general cases--- dark matter particles with and without baryon and lepton charges. (A) Suppose dark matter particles do not carry baryon and lepton charges.  They are still bound to a galaxy by gravity. When the Okubo force pushes the baryon matter, the dark matter in and near a galaxy will move together as a whole.  This resembles the situation where two protons and two neutrons are bound together by the strong nuclear force to form the alpha particle. When the protons are acted on by the electromagnetic force, the neutrons will move together as an alpha particle. (B) Suppose the dark matter particles are, say, neutrinos with very small masses and with lepton charges.  This case is a little bit more complicated because the Okubo-type force can be produced by baryon and lepton charges.  Nevertheless, the galaxy is a gravitationally bound system comprising the stars, other astronomical objects and dark matter, so that they will move together as a whole under the Okubo force.  Based on these simple considerations, it appears that the accelerated expansion of the universe due to the Okubo force will not be qualitatively changed by the presence of dark matter in (and nearby) a galaxy.  Thus, this unified model with the general Yang-Mills symmetry for particle cosmology may be able to provide an understanding of the late-time accelerated expansion of the universe.  A more detailed  discussion of the Okubo force,  dark matter and their possible tests will be given in a separate paper. 
 
It is interesting that, in the unified model for particle-cosmology based on gYM symmetry, the basic force between two point-like baryon charges is constant.  Nevertheless, we have an r-dependent Okubo force between a macroscopic object and a point-like object.
Experimentally, if we are able to measure the frequency shifts of supernovae at different distances from the solar system, we may be able to test qualitatively the prediction (40), even though we do not know where the center of mass of the observable universe is.  In a previous work, we discussed a similar idea closer to the original idea of Lee-Yang concerning the conservation of baryon charge.  It suggests a similar test with accelerated Wu-Doppler shifts.\cite{20}
 
 In this unified model based on the general Yang-Mills symmetry associated with conservation of color charges and baryon charge, an interesting big picture of particle cosmology emerges:  namely, the physics at the smallest scale of quark confinement appears to be intimately related to physics at the largest scale of accelerated cosmic expansion.  Furthermore, the cosmic Okubo force (40) is determined by the basic scale $L_s$ in quark confinement and an effective size of the universe $R_o$, in addition to the extremely small baryon charge $g_b$.
 
{This paper is dedicated to Prof. Susumu Okubo, JP's most respectful mentor.  The author would like to thank D. Fine, L. Hsu and C. L. Yiu for useful discussions.  He also thanks M. R.  Khan for the figure. The work was supported in part by the JingShin Research Fund of the UMassD Foundation.}

\bigskip
\newpage
\noindent
{\large \bf Appendix. \\  Consecutive general Yang-Mills transformations} 
\bigskip

For group properties related to general Yang-Mills symmetry, let us consider
two consecutive general Yang-Mills transformations for the $SU_{N}$ gauge fields, $H_\mu(x)=H_{\mu\nu}^a L^a$,
 \be
 H''_\mu(x) =H'_\mu(x) + \om_{r\mu}(x)  -i [P_r, H'_{\mu}(x)], 
  \ee
   \be
   H'_\mu(x) =H_\mu(x) + \om_{s\mu}(x)  -i [P_s, H_{\mu}(x)],  
   \ee
 $$
 H''_{\mu}=H''^a_{\mu}L^a, \ \ \ \ \ \   \om_{r\mu}=\om_{r\mu}^a L^a.  
$$
 We obtain
\be
H''_\mu=H_\mu + \om_{t\mu} -i[P_t, H_\mu],   
\ee
$$
\om_{t\mu} = \om_{r\mu}+\om_{s\mu},   \ \ \ \ \   P_t = P_r +P_s,
$$
where $P_t$ can be expressed as
\be
  P_t=\left( g \int_{x'_o}^x dx'^\mu \left[\om_{r\mu}^a (x')L^a + \om_{s\mu}^b(x')L^b \right]\right)_{Le(t)},  
\ee 
where $g\equiv g_s$.  To see the equation  $Le(t)$ in (46), let us consider the variation of $P_t$ with a variable end point  $x'_e=x$ and a fixed initial point $x'_o$.  We have
\be
 \de P_t =g  \frac{\p L}{\p\dot{x}^\ld}\de x^{\ld}  + g \left( \int_{\tau_o}^{\tau}\left(-\frac{d}{d\tau}\frac{\p L}{\p \dot{x}^\ld}  
 + \frac{\p L}{\p x^\ld}\right)\de x^\ld d\tau\right)_{Le(t)}
\ee
where we write (46) in the usual form of a Lagrangian with the help of a parameter $\tau$,
\be
P_t=\left( g \int_{\tau_o}^\tau L_t  \ d\tau\right)_{Le(t)}, \ \ \ \ \    L_t=\dot{x}^\mu \om_{t\mu}^a (x) L^a.
\ee

We require that the paths in (47) are those satisfy the Lagrange equation $Le(t)$, i.e., 
\be
-\frac{d}{d\tau}\frac{\p L_t}{\p \dot{x}^\ld} + \frac{\p L_t}{\p x^\ld}=0.
\ee
Thus, the integral in (47) vanishes\cite{8} and we have the relation
\be
 \p_{\mu} P_t=\om^a_{r \mu}(x) L^a+\om^b_{s\mu}(x) L^b.
\ee

The $SU_{N} $ gauge covariant derivatives are  defined as
\be
 \De_{\mu} = \p_{\mu} + ig{H_{\mu}^{ a}} L^a.  
\ee
The $SU_{N}$ gauge curvatures $H_{\mu\nu}=H^a_{\mu\nu}L^a$ are given by
\be
H_{\mu\nu}=\p_\mu H_\nu - \p_\nu H_\mu + ig[H_\mu, H_\nu].
\ee
Under the general Yang-Mills transformations (43) and (44), we have
$$
H''_{\mu\nu}=\p_\mu(H'_\nu+\om_{r\nu}-i[P_r,H'_\nu]) - \p_\nu(H'_\mu+\om_{r\mu}-i[P_r,H'_\mu]) 
$$
$$
+ig[(H'_\mu+\om_{r\mu} -ig[P_r, H'_\mu]),(H'_\nu +\om_{r\nu} -ig[P_r, H'_nu]) ]
$$
$$
=\p_\mu(H_\nu+\om_{t\nu}-i[P_t,H_\nu]) - \p_\nu(H_\mu+\om_{t\mu}-i[P_t,H_\mu]) 
$$
\be
+ig[(H_\mu+\om_{t\mu} -ig[P_t, H_\mu]),(H_\nu +\om_{t\nu} -ig[P_t, H_\nu])].
\ee
After cancellations of the terms $[H_\mu, \om_{t\nu}]$ and $[H_\nu, \om_{t\mu}]$, we obtain
\be
H''_{\mu\nu} +\p_\mu\om_{t\nu} - \p_\nu \om_{t\mu} -ig[P_t, \p_\mu H_\nu - \p_\nu H_\mu]
\ee
$$
-(ig)^2\left( [H_\mu, [P_t, H_\nu]] +  [H_\nu, [H_\nu, P_t]]\right)
$$
$$
=H_{\mu\nu}-ig[P_t, H_{\mu\nu}] +\p_\mu \om_{t\nu} - \p_\nu \om_{t\mu}.
$$

In contrast to the usual gauge symmetry, the $SU_N$ gauge curvatures $H_{\mu\nu}$ do not transform according to the adjoint representation under two consecutive gYM transformations.  However, $\p^\mu H_{\mu\nu}$ has a proper transformation property,
\be
\p^\mu H''_{\mu\nu} =\p^\mu H_{\mu\nu}-ig[P_t, \p^\mu H_{\mu\nu}],
\ee
provided the arbitrary infinitesimal vector gauge functions $\om_{t\mu}$  satisfy the constraints
\be
\p^\mu(\p_\mu \om_{t\nu} - \p_\nu \om_{t\mu}) -ig[\om^\mu_t, H_{\mu\nu}]=0,  
\ee
where $\om_{t\mu}=\om_{r\mu} +\om_{s\mu}$.  This result in two consecutive $SU_N$ gauge transformations being consistent with the constraint~\cite{10} in   (27) for the gYM transformations within the framework of general Yang-Mills symmetry. 

\end{multicols}

\clearpage
\newpage

{\bf \Large Erratum and Addendum}


\begin{multicols}{2}

\bigskip
 `Quark Confinement, New Cosmic Expansion and General Yang-Mills Symmetry' (Hsu J P. Chin. Phys. C. 2017 {\bf 41} (1): 015101)
\bigskip

The force between a gigantic sphere with baryon galaxies exerted on an idealized point-like supernova is given by $F_{cbf}\equiv F_{I}$ in eq. (40) of the original paper.  We can generalized the point-like supernova to a sphere with a radius $R_s$ and a constant mass density $\rh_s$.  Similar to the previous calculations, the modified effective force between the gigantic sphere and a supernova sphere is found to be 
\be
F_{cbf} = \left(\frac{(3g_b)^2 m M}{8\pi L^2_s m^2_p}\right)\left[\frac{r}{R_o}-\frac{r^3}{5 R_o^3} -\frac{ r R^3_s}{5 R^3_o}\right] \equiv F_{II},  
\ee
where $
 m={4\pi R_s^3\rh_s}/{3}. $
In this calculation, one can consider the total baryon charge of the gigantic sphere to be concentrated at the center of the sphere, as indicated by the result of the previous  calculation.  As we expected,  the effective force (3) reduces to that in (40) of the paper in the limit $R_s \to 0$.  

Suppose we use $F_{III}=constant$ to denote the r-independent force for  two point-like baryonic charges.  Our results show that there is qualitative difference between $F_{III}$ and $F_{II}$ in (3).  However, there is no qualitative difference between $F_{II}$ and $F_I$.

This new r-dependent  `Okubo force' (3) for two big baryonic systems is the logical result of the generalization of the original Lee-Yang $U_1$ gauge symmetry for conserved baryonic charges.\cite{1}    Such a `Lee-Yang symmetry' is associated with the usual baryonic gauge field and the inverse-square baryonic force.  Based on E$\ddot{o}$tv$\ddot{o}$s experiment, they estimated that their baryonic force is about $10^{-5}$ times weaker than the gravitational force.   Thus, both the `Lee-Yang baryonic force' and the `Okubo force' (as estimated in (42) of the paper) are extremely weak in comparison with the gravitational force.  However, the repulsive
 Okubo force (3) among baryon galaxies can overcome the gravitational attractive force when the distance is large enough.  Thus, the Okubo force (3) can provide a gauge-field-theoretic understanding of the discovery of the Hubble Space Telescope.  Namely, there was cosmic deceleration that preceded the current epoch of cosmic acceleration.  These calculations can be carried out in a framework of spacetime with a broad\cite{2} four-dimensional symmetry.

\end{multicols}



\bigskip

\begin{thebibliography}{90}

\vspace{3mm}

\bibitem{1}LEE T D,  Particle Physics and Introduction to Field Theory. Harwood Academic Publishers, 1981. ch. 10.  

\bibitem{2}HSU J P,  Mod. Phys. Letts. A,  2005, {\bf 20}: 2855 

\bibitem{3}Olive K A et al. (Particle Data Group), Chinese Phys. C, 2014 {\bf 38}: 09000,  (http://pdg.lbl.gov).  

\bibitem{4}HSU J P, Cottrell K O, Chinese Phys. C. 2015, {\bf 39}: (10) 105101

\bibitem{5}HSU J P, Eur. Phys. J. Plus, 2014 {\bf 129}: 108 

\bibitem{6} Yourgrau W, Mandelstam S, Variational Principles in Dynamics and Quantum Theory. Dover, 3rd ed, 1979.  50.

\bibitem{7}Pais A and Uhlenbeck G E, Phys. Rev. 1950, {\bf 79}: 145 

\bibitem{8}Landau L, Lifshitz E, The Classical Theory of Fields. Addison-Wesley, 1951. 29 

\bibitem{9}HUANG K,  Quarks, Leptons and Gauge Fields.  Singapore: World Scientific, 1982.  241-249. 

\bibitem{10}Lee B W, Zinn-Justin J, Phys. Rev., 1973, {\bf 7}: 1047   

\bibitem{11}Eichten E, Gottfried K, Kinoshita T,  Kogut J, Lane K D, Yan T -Y, Phys. Rev. Lett., 1975, {\bf 34}: 369 

\bibitem{12}Gel'fand I M, Shilov G E, {\em Generalized Functions}
vol. 1, New York: Academic Press, 1964. 363.

\bibitem{13}Fowles G R, Cassiday G L, Analytical Mechanics. Thomson, 2005. 223-225

\bibitem{14}CHOU K C, WU Y L, Phys. Rev. D, 1996, {\bf 53}: R3492

\bibitem{15}Weinberg S, The Quantum Theory of Fields Vol. 2. London: Cambridge Univ. Press, 1996. 152-157

\bibitem{16}HSU J P, HSU L, Space-Time Symmetry and Quantum Yang-Mills Gravity. Singapore: World Scientific, 2013.  214-215, 225

\bibitem{17}HSU L and HSU J P, In: Proc. of the Twelfth Asia-Pacific Int. Conf. on Gravitation and Astrophysics. Singapore: World Scientific, 2016.  eds. Melnikov V and HSU J P.  3, 369

\bibitem{18}XIA T Y,  ZHANG Y, Phys. Lett. B 2007 {\bf 656}: 19  

\bibitem{19}See, for example, Okabe N, et al,   Astrophys. J. Letts, 2013 {\bf 769} ID.35  

\bibitem{20}HSU J P, NING Z H, In: Gravitation and Astrophysics, Proc. of the VII Asia-Pacific Int. Conf., Singapore: World Scientific, 2007.  eds. Nester J M, CHEN C M, HSU J P.  79-86

\end{thebibliography}

\bigskip

\begin{thebibliography}{90}

\vspace{0mm}



\bibitem{1}Lee T D, and Yang C N. Phys. Rev., 1955, {\bf 98}, 1501

\bibitem{2}Hsu J P, and Hsu L. Phys. Letters A, 1994 {\bf 196}, 1




\end{thebibliography}
\end{document}